\documentclass[a4paper,fleqn,usenatbib]{mnras}

\usepackage[T1]{fontenc}
\usepackage{ae,aecompl}
\usepackage{float}
\usepackage{cleveref}
\crefname{section}{§}{§§}
\Crefname{section}{§}{§§}

\usepackage{graphicx}	
\usepackage{amsmath}	
\usepackage{amssymb}	

\title[The Mass Density Profiles and Star Formation History]{The Mass Density Profile and Star Formation History  of Gaussian and Non-Gaussian Clusters}

\author[Reinaldo R. de Carvalho]{
R. R. de Carvalho$^{1,4}$\thanks{E-mail: rrdecarvalho2008@gmail.com},
A. P. Costa$^{2}$\thanks{E-mail: apcosta@uesc.br},
T. C. Moura$^{3}$\thanks{E-mail: tatiana.mourabastos@gmail.com} and
A. L. B. Ribeiro$^{2}$\thanks{E-mail: albr@uesc.br} 
\\
$^{1}$NAT - Universidade Cruzeiro do Sul / Universidade Cidade de S\~ao Paulo\\
$^{2}$Laborat\'orio de Astrof\'{\i}sica Te\'orica e Observacional, Universidade Estadual de Santa Cruz - 45650-000, Ilh\'eus-BA, Brazil\\
$^{3}$Instituto Astrof\'isico e Geof\'isico da USP, S\~ao Paulo, 05508-090, SP, Brazil\\
$^{4}$Divis\~ao de Astrof\'{\i}sica (INPE-MCT), S\~ao Jos\'e dos Campos, 12227-010, SP, Brazil
}

\date{Accepted XXX. Received YYY; in original form ZZZ}

\pubyear{2017}

\begin{document}
\label{firstpage}
\pagerange{\pageref{firstpage}--\pageref{lastpage}}
\maketitle

\begin{abstract}

This paper is the third of a series in which we investigate the discrimination between Gaussian (G) and Non-Gaussian (NG) clusters, based on the velocity distribution of the member galaxies. We study a sample of 177 groups from the Yang catalog in the redshift interval of 0.03 $\le$ z $\le$ 0.1 and masses $\ge$ 10$^{14} \rm M_{\odot}$.  Examining the projected stellar mass density distributions of G and NG groups we find strong evidence of a higher infall rate in the outskirts of NG groups over the G ones. There is a 61\% excess of faint galaxies in NGs when contrasted with G groups, when integrating $\rm from ~ 0.8 ~to~ 2.0R/R_{200}$. The study of the Star Formation History (SFH) of ellipticals and spirals in the three main regions of the Projected Phase Space (PPS) reveals also that the star formation in faint spirals of NG groups is significantly different from their counterpart in the G groups. The assembled mass for Faint spirals varies from 59\% at 12.7 Gyr to 75\% at 8.0 Gyr, while in G systems this variation is from 82\% to 91\%. This finding may also be interpreted as a higher infall rate of gas rich systems in NG groups. This accretion process through the filaments, disturbing the velocity distribution and modifying not only the stellar population of the incoming galaxies but also their SFH, should be seriously considered in modelling galaxy evolution.

\end{abstract}

\begin{keywords}
Galaxies -- Galaxy Cluster -- method: Star Formation History -- Mass Density Distribution.
\end{keywords}

\section{Introduction}

In the hierarchical structure formation framework ($\Lambda$CDM), clusters of galaxies grow from higher density regions in the large-scale structure of the early universe. In these regions, the galaxy formation process is likely to be accelerated, which explains qualitatively why the majority of star formation in galaxies located in the cluster central regions takes place much earlier ($z\gtrsim1$) than that in the general field (e.g. \citealt{1998ApJ...504L..17V}, \citealt{2000ApJ...531..184K}, \citealt{2003MNRAS.346....1K}). The cluster formation is a continuous process and clusters may be found in different dynamical stages, as indicated by recent studies (\citealt{2009ApJ...702.1199H}; \citealt{ribeiro2013spider}; \citealt{2017MNRAS.467.3268R}; \citealt[][Paper I]{1538-3881-154-3-96}; \citealt[][Paper II]{2018MNRAS.473L..31C}). Hence, to understand the physical processes that drive galaxy evolution, it is imperative to analyse the properties of galaxies in different environments. Defining environment is no trivial task and it may be approached in different ways: local density, distance from cluster center, halo mass and, as more recently suggested, through the examination of the velocity distribution of galaxies in clusters, G (Gaussian) and NG (Non Gaussian) systems \citep{1538-3881-154-3-96}. In the later case there is an important assumption made regarding the dynamics of the gravitational system - G systems are supposed to be in equilibrium \citep[e.g.][]{Ogorodnikov, Lynden-Bell67}.

An important parameter used to investigate the relation between galaxy properties and the environment in which they inhabit is the stellar mass, which provides an evolutionary trace of the galaxy population in terms of star formation rate and morphology (e.g. \citealt{2007ApJ...663..834B}, \citealt{2008ApJ...686...72C}, \citealt{2010MNRAS.406..147F}), and it is also a possible discriminator of the distribution of the underlying dark matter in galaxy systems, since stars are not expected to be stripped significantly because they are located deep in the galaxy potential and both, stars and dark matter, are collisionless (e.g. \citealt{2015A&A...577A..19V}, \citealt{2015A&A...575A.108A}). In this sense, \cite{2010MNRAS.402.1796W} studying galaxy clustering in the Millennium Simulation, (\citealt{2005Natur.435..629S}) proposed an empirical method to link galaxy stellar mass directly with its hosting dark matter halo mass. In that work, while the positions of galaxies are predicted by following the merging histories of halos and the trajectories of sub-haloes in the Millennium Simulation, the stellar mass of galaxies at $z\sim0$ is related to the mass of the halo at the time when the galaxy was last the central dominant object.

All these studies relating the degree of gaussianity of the velocity distribution to the properties of galaxies in groups/clusters (\citealt{ribeiro2013spider}; \citealt{1538-3881-154-3-96}; \citealt{2018MNRAS.473L..31C}) have proved the importance of assessing the velocity distribution, or even better the PPS, to gain more physical understanding of the processes operating in galaxies (\citealt{Biviano2002}; \citealt{Muzzin2014}; \citealt{Jaffe18}). In this Letter, we focus our attention on the stellar mass distribution and the SFH of galaxies located in 177 groups selected by de Carvalho et al. (2017) from the Yang catalog (Yang et al. 2007). We describe the sample and methodology used in Section 2. The results obtained for the stellar mass distribution are presented in Section 3 followed by the analysis of the star formation history in Section 4. Section 5 discusses the main findings of this contribution. We adopt a cosmology with H$_{\circ} = 72 ~\rm km/s/Mpc$, $\Omega_{m}$ = 0.3, and $\Omega_{\Lambda}$ = 0.7.
 
\section{Data and Methodology}

In this section, we briefly describe the sample, data and the procedures used to identify the dynamical state of the groups. In the following analysis, we use 177 groups from the Yang group catalog representing systems with halo mass greater than 10$^{14.0} \rm\hspace{1mm} M_{\odot}$, and classified as G or NG by \cite{1538-3881-154-3-96}. The gaussianity of the velocity distribution of galaxies in clusters is measured using the so-called Hellinger Distance, a parameter that measures the distance between a given distribution and a gaussian (see \citealt{1538-3881-154-3-96} for more details). The sample consists of 5,068 galaxies in G groups and  2,987 galaxies in NG's. We define the Bright domain as ($0.03\leq z \leq 0.1$ and $M_{r}\leq -20.55$, $\sim M^{\star}+1$), where $M_{r}\leq -20.55$ is the limiting absolute magnitude, in r-band, corresponding to the spectroscopic completeness of SDSS-DR7 at z = 0.1, while Faint means  ($0.03\leq z \leq 0.04$ and $-20.55 < M_{r} \leq -18.44$). When we refer to the Faint domain this includes only galaxies in groups in the redshift bin of  $0.03 \leq \rm z \leq 0.04$ probing the luminosity function down to $\sim \rm M^{\star}+3$. The parameters used in this investigation are presented in \cite{1538-3881-154-3-96}. Halo mass, M$_{200}$, and virial radius, R$_{200}$, were estimated using the shift-gapper analysis described in \cite {Lopes2009}. Stellar masses were determined after characterizing the stellar population using the spectral fitting code Starlight \citep{CMSS05}. The stellar masses are determined within the fiber aperture and the extrapolation to the full extent of the galaxy is done by calculating the difference between fiber and model magnitudes in the z band (see \citealt{1538-3881-154-3-96}).

As far as morphological classification, we have used the catalog provided by the second edition of the Galaxy Zoo Project (hereafter Zoo2). This catalog counts 243,500 bright, near and large galaxies with available spectroscopic data in the seventh release of the SDSS database. We have  $\sim$ 80\% (4060/5068) of the galaxies in G Groups and $\sim$ 78\% (2344/2987) of the galaxies in NG Groups with morphological classes given by Zoo2. Finally, three different regions of the PPS are defined: Virial (VIR) - $0.0<{\rm R/R_{200}}<0.5$ and $|\Delta V/\sigma|<1.0$; BackSplash (BS) - $1.0<{\rm R/R_{200}}<1.5$ and $|\Delta V/\sigma|<1.0$; and Infall (INF) - $1.5<{\rm R/R_{200}}<2.0$ and $1.0<|\Delta V/\sigma|<2.0$. 

\section{Stellar Mass Distributions}

Although not much is known about the stellar mass distribution in clusters (e.g. \citealt{2013A&A...550A..58V}), recent works (e.g. \citealt{2009ApJ...703..982G}, \citealt{2014A&A...561A..79V}, \citealt {2014A&A...571A..80A} \citealt{2015A&A...577A..19V}, \citealt{2016A&A...585A.160A}) have been shedding some light into the properties of galaxies and their host environment by measuring the stellar mass density profiles $\Sigma_{M}$ (the stellar mass included in each radial ring divided by its area). In this work, we use the stellar mass density profiles to investigate how galaxies are distributed in G and NG systems.

\begin{figure}
\hspace{-0.3in}
{\hspace{0.05in}\includegraphics[height=8.8cm, width=8.8cm]{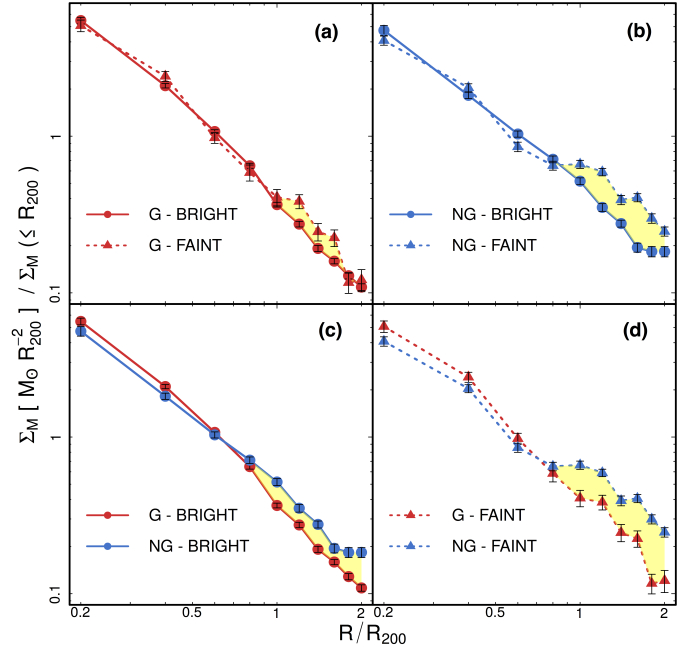}}
\caption{Projected surface stellar mass density profiles for the Yang clusters in our sample. In panel (a) we compare the Bright and Faint components of G groups, while in panel (b) we show how these two components behave in NG groups. Panel (c) exhibit the comparison between the Bright component of Gs and NGs, and panel (d) compares the Faint component of Gs and NGs. Error bars are displayed as the standard deviation, obtained by performing 1000 
bootstraps of the stellar mass values in each radial ring. Notice that sometimes these error bars are smaller than the size of the symbols representing our estimates.}
\label{fig01}
\end{figure}

In Figure \ref{fig01}, we show our stellar mass density profiles. Before we stack all systems belonging to each dynamic group, we normalize the projected radial distances by $R_{200}$, which ensures not to mix  inner regions, virialized, with outer regions, non-virialized. The stellar mass densities were scaled by the stellar mass density corresponding to the virialized region ($\leq R_{200}$) for similar reasons to those considered in the normalization of the projected distances.

A common practice in this type of study is fitting the stellar mass density profiles with an NFW model (\citealt{1997ApJ...490..493N}) to measure the total stellar mass of the system. However, this procedure brings an operational problem due to the fact that the stacked profile represents the mean distribution, disregarding the peculiarities of each system, and the amount of stellar mass, possibly significant, in-falling into the cluster. Since the NFW profile model does not consider the complexity of the data being fitted, in this work we choose, similarly to \cite{2015A&A...575A.108A}, not to perform this fitting. An example of the deviations of the NFW model from the surface mass density profiles can be found in \cite{2014ApJ...795..163U}.

In Figure \ref{fig01}, we show the stacked stellar mass density profile for the Bright and Faint components of the G and NG groups. In all panels we see a clear excess of one component over the other (yellow shaded area). We quantify this excess by doing numerical integration under the curves between two specific radius as indicated below ($\int {\Sigma_{M} / \Sigma_{M}(\leq R_{200}) ~ dR}$). We note that in stacking all the stellar mass density profiles of clusters in G and NGs, we normalize these stellar mass densities by the stellar mass density within the virialized region. This procedure implies loosing the proper stellar mass density scale, which means that we are measuring only relative excesses.

In panel (a), we compare the Faint and the Bright components of G groups and we do observe a small excess of the Faint over the Bright ($\sim$24\%) when we integrate from $1.0 ~ \rm to ~2.0R/R_{200}$. For NG systems (panel b), we see the same behavior, although the excess of the Faint over the Bright component in the outer region ($0.8~ \rm to ~2.0R/R_{200}$) is more pronounced (42 \%). This might indicate a higher infall rate of low mass galaxies in NG systems compared to Gs. From panel (c), we learn that the Bright component in NG systems is also in clear excess (30\%) in the outskirts ($\rm from ~ 0.8 ~to~ 2.0R/R_{200}$) over the G groups. The same trend is present when comparing the Faint components of Gs and NGs (panel d) - the Faint component in NGs exhibits a considerable excess (61\%) in surface stellar mass density than their counterparts in Gs.


\begin{figure} 

{\hspace{-0.2in} \includegraphics[width=8.8cm, height=8.8cm]{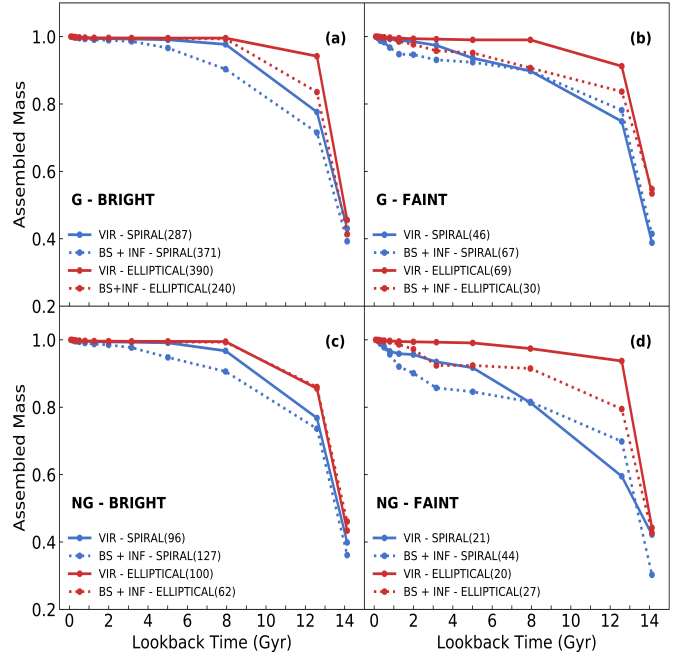}}
\caption{Fraction of assembled stellar mass as a function of the lookback time in Gyr, determined directly from
the base ages. Solid red line represents ellipticals in the VIR region; Dotted red line means ellipticals in the BS+INF
region. Symbols are the same for spirals but in blue. The gaussianity as well as the luminosity regime probed are indicated in each panel. The number of galaxies in each morphological type and locus on the PPS are listed accordingly.}
\label{fig02}
\end{figure}

        
\section{Star Formation History}

Our spectral analysis results not only in averaged age and metallicity but also in the SFH, which for a given galaxy is expressed by the contribution, as mass fraction, from each basis SSP (Single Stellar Population). We estimate the cumulative fraction of stars older than a given age, as a function of the age of the basis. Therefore, the SFH is measured as the cumulative distributions over all galaxies in the sample within the two luminosity regimes, Bright and Faint. These luminosity regimes translate into two stellar mass domains, which here we list as the quartiles and median of the distributions (Q$_{25\%}$, Median, Q$_{75\%}$) - for the Bright domain we have (10.30, 10.49, 10.65) and for the Faint (9.26, 9.61, 9.89). Figure \ref{fig02} shows the SFH, expressed as the assembled mass versus the lookback time (LBT). The accuracy of the SFH is directly related to the age measurement. To assess how well we measure it, we use two different approaches: 1) Subsampling - where the resample size (70\%) is smaller than the sample size and resampling is done without replacement (1000 realizations); 2) Using repeated observations, which is dominated by bright galaxies.  We use all galaxies in the sample with galaxy zoo classification, 4442 in the Bright regime and 872 in the Faint one. As for using the repeated observations, we have 2698 galaxies (6148 spectra). We quote the accuracy in SFH as Q$_{\sigma}$ = 0.7415 (Q$_{75\%}$ - Q$_{25\%}$). Using the repeated observations we find that up to 3 Gyr the error is $\sim$1\%, from 3 to 8 Gyr is $\sim$6\%, reaching $\sim$14\% for ages larger than 10 Gyr. For faint galaxies (lower S/N) the errors are larger, up to 3 Gyr the error is $\sim$5\%, from 3 to 8 Gyr is $\sim$13\%, reaching $\sim$22\% for ages larger than 10 Gyr. Both methods of evaluating errors provide similar results. 

In panel (a) of Figure \ref{fig02}, bright galaxies in G systems, we see that ellipticals (in red) and spirals (in blue) have the same overall behavior in assembling stellar mass {\it wrt} the environment. Galaxies in the Virial region (solid line) tend to form stars earlier and faster than their counterpart in the general infall region (dotted line). We also see that ellipticals and spirals, regardless of their position in the PPS, complete forming all their stars in the last 3 Gyr, except the spirals in the VIR and BS+INF regions of NG systems (panel d). Notice that at early stages (LBT$\sim$ 13 Gyr) spirals have assembled only $\sim$75\% of their mass while ellipticals already assembled at least $\sim$ 84\%, depending where they are located in the PPS. Ellipticals in the VIR region can accrete almost 95\% of their mass at LBT$\sim$ 13 Gyr (monolithic scenario). Spirals, on the contrary, have a more gradual star formation history, with those in the BS+INF region accreting 95\% of their mass only at LBT$\sim$ 6 Gyr. This is the picture that emerges when focusing on the bright galaxies of G clusters, which is in agreement with results presented in \cite{2012ApJ...752L..27T}. We exhibit the SFH for faint galaxies in G systems in panel (b) of Figure \ref{fig02}. The comparison with the bright galaxies reveals important aspects of the way galaxies evolve in relaxed systems. The faint ellipticals have their star formation history very similar to the bright ones, although the ones in the infall region seem to have a more extended star formation history (dotted red line). Also, faint spirals have the same extended SFH, and again we see a more extended star formation for the ones in the BS+INF region. Notice also that ellipticals in the infall region look like spirals in the VIR+BS+INF regions. This may be partly reflecting our inability to properly classify galaxies when they are in the faint regime, namely spirals in VIR+BS+INF and ellipticals in the BS+INF represent systems that gradually form their stars until recently.

In panel (c) of Figure \ref{fig02}, bright galaxies in NG systems, we note that ellipticals have the same SFH regardless of the environment, namely same behavior in the VIR, BS and INF regions, and similar to what we find for the bright galaxies in G systems. This result indicates that the dynamics of massive galaxies in NG systems is not significantly affected by the infall from the group surroundings. The same is true for the SFH for spirals, there seems to be no difference when comparing how star formation proceeds in spirals located in G or NG groups. In summary, we find that bright galaxies exhibit a similar SFH in G and NG systems, which reinforces previous findings for the Age and metallicity distributions of these galaxies when comparing also G and NG groups \citep[see][]{1538-3881-154-3-96}. Finally, we examine the SFH of faint galaxies in NG groups and here the differences between G and NG are very significant. Faint ellipticals in NG systems have their SFH similar to their counterparts in G systems. On the other hand, Faint spirals in NG groups, display a very different behavior in comparison with the G ones. In the BS+INF region, they accrete around 72\% of their mass at very early stages and then gradually form stars until recently (LBT$\sim$ 3 Gyr). They assemble $\sim$85\% of their mass and then experience a sudden growth of their stellar mass from $\sim$85\% to completion. In the VIR region they have a steeper behavior in SFH from early times up to ~ 5 Gyr when then they evolve more slowly.  These striking differences reinforce the differences between G and NG groups specially when we examine the faint galaxies in the outskirts (e.g. \citealt{1538-3881-154-3-96}, \citealt{2018MNRAS.473L..31C}).

The overall picture that emerges from Figure \ref{fig02}, when looking at the bright galaxies, is in agreement with previous works (e.g. \citealt{2008ApJ...675..234P},  \citealt{2011MNRAS.418L..74D}, \citealt{2012ApJ...752L..27T}). Ellipticals in the VIR region, assemble more than 95\% of their mass at LBT $\sim$ 10 Gyr. Spirals in general form stars more gradually, exhausting the gas component only recently  \cite{2008ApJ...675..234P}. An important caveat in this analysis is that when we set a given galaxy to a specific position in the PPS, due to the statistical nature of this association, we introduce a certain variance in the SFH. The fraction of virialized, infall and backsplash galaxies in cells of PPS, as listed by \cite{MMR11}, represent the probability of recovering a galaxy properly associated to the corresponding cells, namely virial (89\%), backsplash (35\%) and infall (97\%).

\section{Discussion}

Examining a sample of 319 groups from the Yang's catalog (\citealt{Yang07}), \cite{1538-3881-154-3-96} find that 76\% of the Yang's groups with masses above 10$^{14} \rm M\odot$ have Gaussian velocity distributions. Estimating skewness and kurtosis of the velocity distribution, they find evidence that faint galaxies in the outer regions of NG groups are infalling for the first time into the systems. Also, in the inner regions of G groups, galaxies are older and more metal-rich than their counterparts in NG systems. As far as the outer regions are concerned, galaxies of NG groups are older and more metal-rich than galaxies in G's, independent of their luminosity. These are clear indications of preprocessing in the outer parts of NG groups \citep{Fujita04}, in agreement with \cite{2017MNRAS.467.3268R}. \cite{2018MNRAS.473L..31C} study the velocity dispersion profiles (VDP) of G and NG systems and find that the stacked VDP for G groups exhibit a central peak followed by a monotonically decreasing behavior indicating predominantly radial orbits. The Bright stacked VDPs show lower velocity dispersions than the Faint stacked VDPs, in all radii. As far as NG systems, they display a distinct feature with a depression in the central region and also a likely higher infall rate associated with galaxies in the Faint stacked VDP. These findings corroborate the \cite{2017A&A...606A.108C} results for regular and irregular systems.  We note that what \cite{2017A&A...606A.108C} find for VDPs for regular systems are similar to that of a G system and the VDP for S galaxies in the irregular systems exhibit a depression in inner regions, similar to what \cite{2018MNRAS.473L..31C} find for the NG systems.

In this contribution, we investigate how the stellar mass distribution and the SFH of galaxies in clusters depend on the dynamical stage of groups/clusters. In \cite{1538-3881-154-3-96}, we present a robust technique to measure how the velocity distribution of galaxies is far from gaussianity, and how that reflects the physical characterization of an out of equilibrium system. This distinction between G and NG systems is the base of our attempt to characterize the link between environment and galaxy properties.

We notice that, as pointed out by \cite{2009MNRAS.400..937M} in their semi-analytic models, massive clusters ($>10^{14.5}\rm M_{\odot}$) at redshift zero have accreted 40\% of their galaxy members from the external environment. This gives further support to the idea presented here that it may be hard to find nearby clusters with perfect gaussian velocity distribution and that gaussianity, as measured in this paper, reflects the current infall rate. The projected surface stellar mass density profiles exhibit clear evidence of a higher infall rate between 1 and 2 $\rm R_{200}$ in NG groups compared to the G ones. The behavior displayed in panels (a) and (b), with a clear excess of the Faint component over the Bright one, is in agreement with results presented by \cite{Sanchez08} who find an increase of the dwarf-to-giant ratio towards the outskirts of clusters. However, such an increase seems larger for NG systems considering that the excess of the Faint component over the Bright one for NG groups (42\%) is about twice as large as the excess measured for G systems (24\%). This result reinforces previous findings of \citealt{1538-3881-154-3-96} and \citealt{2018MNRAS.473L..31C}), namely, a higher infall rate in NG groups disturbs the mean stellar population of galaxies and the velocity dispersion distribution of galaxies in the outskirts of these systems.

In recent years, a lot of attention has been paid to the question of what drives the evolution of galaxies in the outskirts of groups and clusters \citep{Boselli16, Jaffe18} . A galaxy in the vicinity of a rich cluster starts interacting gravitationally with other galaxies more frequently and also it senses the potential well of the system as a whole \citep{Moore98}. This may alter the stellar population properties of these infalling galaxies. Therefore, we do expect significant variations of their SFH specially when we probe the PPS for different morphological types. In this work, we find that bright ellipticals in the VIR region (panels a and c of Figure 2), assemble more than 95\% of their mass at LBT $\sim$ 10 Gyr, regardless if they are in G or NG systems, which is a somewhat expected result \citep{2012ApJ...752L..27T}. Bright spirals assemble their stellar mass in a similar way comparing G and NG groups, independent of the PPS region. As far as Faint galaxies go, we see a very different behavior. Faint ellipticals have similar SFH in G and NG groups and they are equally old in both systems \citep{1538-3881-154-3-96, Einasto18}. However, the SFH of Faint spirals in NG systems is considerably dissimilar to those in G groups. The SFH of Faint spirals in the VIR region of NG groups grows more slowly than their counterpart in G groups. This can be interpreted as further evidence of a higher infall rate in NG systems. 

In summary: bright galaxies (ellipticals and spirals) seem to have similar SFH independently of where they are located in the PPS; faint ellipticals also have similar SFH regardless the location in the PPS and the key difference arises when  we examine faint spirals in the VIR and BS+INF regions - SFH of galaxies in the VIR region belonging to NG groups is far more gradual. These conclusions are critically important considering that if the distinction between G and NG is not made, this can imply a potentially serious problem in modelling galaxy evolution in clusters.


\section*{Acknowledgements}

RRdC thanks Francesco La Barbera, Gary Mamon, Joseph Silk and Tatiana Lagana for fruitful discussions on this topic. APC thanks CAPES financial support, ALBR thanks CNPq, grant 311932/2017-7 and RRdC acknowledges financial support from FAPESP through grant $\#$2014/11156-4



\bibliographystyle{mnras}
\bibliography{references.bib} 

\bsp	
\label{lastpage}
\end{document}